\documentclass[review]{elsarticle}

\usepackage{lineno}
\usepackage[hyphens]{url}
\usepackage[colorlinks=true, allcolors=blue]{hyperref}
\usepackage[draft]{fixme}
\modulolinenumbers[5]
\usepackage{xcolor}
\usepackage{subcaption}
\usepackage{soul}
\usepackage{todonotes}

\usepackage{listings}
\usepackage{xcolor}

\journal{Journal of \LaTeX\ Templates}

\begin{document}

\begin{frontmatter}

\title{Deployment of Elastic Virtual Hybrid Clusters Across Cloud Sites}





\author[I3M]{Miguel Caballer}
\author[INFN]{Marica Antonacci}
\author[CESNET]{Zdeněk Šustr}
\author[UNIBA,INFN]{Michele Perniola}
\author[I3M]{Germ\'an Molt\'o\corref{mycorrespondingauthor}}
\ead{gmolto@dsic.upv.es}

\address[I3M]{Instituto de Instrumentaci\'on para Imagen Molecular (I3M)\\
Centro mixto CSIC - Universitat Polit\`ecnica de Val\`encia \\
Camino de Vera s/n, 46022, Valencia
} 

\address[INFN]{INFN Bari, Via E. Orabona 4, 70125 Bari, Italy}

\address[CESNET]{CESNET, Zikova 4, 160\,00 Praha 6, Czechia}

\address[UNIBA]{Department of Physics, University of Bari, Via G. Amendola 173, 70126 Bari, Italy}
	
\cortext[mycorrespondingauthor]{Corresponding author}

\begin{abstract}
Virtual clusters are widely used computing platforms than can be deployed in multiple cloud platforms. The ability to dynamically grow and shrink the number of nodes has paved the way for customised elastic computing both for High Performance Computing and High Throughput Computing workloads. However, elasticity is typically restricted to a single cloud site, thus hindering the ability to provision computational resources from multiple geographically distributed cloud sites. To this aim, this paper introduces an architecture of open-source components that coherently deploy a virtual elastic cluster across multiple cloud sites to perform large-scale computing. These hybrid virtual elastic clusters are automatically deployed and configured using an Infrastructure as Code (IaC) approach on a distributed hybrid testbed that spans different organizations, including on-premises and public clouds, supporting automated tunneling of communications across the cluster nodes with advanced VPN topologies. The results indicate that cluster-based computing of embarrassingly parallel jobs can benefit from hybrid virtual clusters that aggregate computing resources from multiple cloud back-ends and bring them together into a dedicated, albeit virtual network.

\end{abstract}

\begin{keyword}
Cloud Computing, Network virtualization, Cluster computing
\end{keyword}

\end{frontmatter}


\section{Introduction}
Scientific computing typically requires execution of resource-intensive applications that require collaborative usage of multiple computing and storage resources in order to satisfy the demands of the applications. This is why Distributed Computing Infrastructures (DCIs), such as the European Grid Infrastructure (EGI) \cite{Kranzlmuller2010} or the Open Science Grid (OSG) \cite{Altunay2011}, emerged in the last decades to encompass the rise of eScience. Indeed, as described in  \cite{Medeiros2016}, ``eScience studies, enacts, and improves the ongoing process of innovation in computationally-intensive or data-intensive research methods; typically this is carried out collaboratively, often using distributed infrastructures''.

The advent of cloud computing \cite{NISTCloud}, exemplified by public cloud providers such as Amazon Web Services, Microsoft Azure or Google Cloud Platform, together with Cloud Management Platforms such as OpenStack and OpenNebula, popularised access to on-demand elastic computing, storage and networking capabilities. This represented a major step forward from previous Grid infrastructures, in which users had to adapt their applications to run on the Grid. Instead, using cloud technologies, users can customize the virtualized execution environment to satisfy the requirements of the application, not the other way round.

However, coordinated execution on these customized computing environments was still required to execute scientific applications. Indeed, High Performance Computing (HPC) applications, typically based on message-passing libraries such as the Message Passing Interface (MPI) or shared-memory programming libraries, such as OpenMP, require a coordinated execution of a set of processes on the computing infrastructure. To this aim, computing clusters have been widely adopted in the last decades as the preferred computing platform to run scientific applications. 

A computing cluster is a set of nodes typically interconnected by low-latency networks that use a Local Resource Management System (LRMS) together with a shared file system to ease operations across all the nodes of the cluster. Popular examples of LRMS are SLURM \cite{Slurm} or HTCondor \cite{Thain2005a,Fajardo2015}, which schedule jobs to a finite set of computing resources provided by the nodes of the computing cluster.

However, the inherent acquisition and maintenance cost of physical clusters, together with the general availability of cloud providers, made it economically viable for certain workloads to deploy virtual elastic clusters on the cloud. A thorough discussion on the viability of outsourcing cluster computing to a cloud is available in a previous work by the authors \cite{DeAlfonso2013a}.

Furthermore, the coexistence of on-premises and public clouds has leveraged cloud bursting, where virtual clusters can be enlarged with resources outside the organization and, therefore, hybrid clusters can harness on-premises and public cloud resources simultaneously. This approach is highly advantageous because users can seamlessly access cluster-based computing resources in addition to those available via their on-premises clouds. Other topologies can be considered for hybrid clusters when using virtual resources, such as heterogeneous clusters, where various nodes in the cluster have different hardware characteristics.

However, when deploying hybrid virtual elastic clusters across multiple IaaS cloud sites several challenges need to be solved, which include, but are not limited to: i) deciding on the set of cloud sites that are most suitable to be chosen for the hybrid deployment; ii) provisioning and configuring the computing nodes across multiple cloud back-ends; iii) automatically creating secure networking tunnels that provide seamless connectivity among geographically distributed nodes of the cluster; iv) minimising the number of public IPv4 addresses involved in the deployment of the cluster to cope with the restricted quotas that many research centers face; v) taking advantage of level~2 capabilities of user-created private networks in Cloud Management Frameworks such as OpenStack.

To this aim, this paper describes an architecture that integrates the following open-source components: the INDIGO-DataCloud PaaS Orchestrator \cite{Orchestrator} that collects  high-level deployment requests from the software layer, and coordinates the resource or service deployment over dynamic Apache Mesos\footnote{Apache Mesos: \url{http://mesos.apache.org/}} clusters or directly over IaaS platforms using continuous configuration and DevOps-based deployment approaches; the Infrastructure Manager (IM), which deploys complex and customized virtual infrastructures on IaaS cloud deployments, and the INDIGO Virtual Router \cite{vRouter}, to establish overlay virtual private networks to interconnect nodes in a cloud platform deployment even if deployed in multiple, geographically distant sites. The main contribution of this paper is to showcase an open-source integrated platform to provide users with the ability to dynamically provision elastic virtual clusters across hybrid clouds that can seamlessly support cloud bursting supported by flexible virtual routing in hybrid setups to accommodate workload increases.

After the introduction, the remainder of the paper is structured as follows. First, section \ref{sec:related-work} describes the related work in the area of virtual clusters and hybrid networking. Next, section \ref{sec:architecture} introduces the proposed architecture to deploy the virtual elastic clusters on hybrid infrastructures composed of several IaaS clouds. Then, section \ref{sec:use-cases} describes a case study that uses the aforementioned architecture to deploy virtual clusters with automated networking connectivity among the nodes with hybrid resources from both research centers and public clouds. The results are analysed to point out the benefits and limitations of the proposed approach. Finally, section \ref{sec:conclusions} summarises the main achievements of the paper and points to future work.

\section{Related Work \label{sec:related-work}}

There are previous works in the literature that aim at deploying virtual clusters on cloud infrastructures: ElastiCluster \cite{ElastiCluster} is a command line tool to create, manage and setup computing clusters hosted on cloud infrastructures (like Amazon’s Elastic Compute Cloud EC2, Google Compute Engine or OpenStack-based clouds). The clusters must be scaled by the user and, therefore, automated elasticity is not supported. Furthermore, although different cloud providers are supported, hybrid deployments cannot be performed since each cluster must be provisioned within a single provider. StarCluster \cite{StarCluster} is an open-source tool that provisions clusters in Amazon EC2 based on a predefined configuration for applications (Open Grid Scheduler, OpenMPI, Network File System (NFS), etc.). It provides automated elasticity but remains restricted only to AWS resources. Furthermore, this project is no longer under active development. The work by Coulter et al. \cite{Coulter2019} uses SLURM's built-in power-saving mechanisms to add nodes to clusters using scripts to call the cloud provider's API to create/remove nodes. In this case, the cluster solution is bound to a specific batch system and does not support hybrid deployments. The works by Yu et al. \cite{Yu2016, Yu2018} are mainly focused on network virtualization with bandwidth guarantee for virtual clusters, with no mention to multi-site approaches.

Another relevant tool is EC3 (Elastic Cloud Computing Cluster) \cite{Caballer2013}, an open-source tool to deploy virtual elastic clusters on multiple clouds (both public and on-premises) that features a web-portal integrated in the EGI Applications on Demand \cite{Sipos2019} portfolio of solutions for the long tail of science. EC3 was extended in previous works to support hybrid deployments, i.e., the ability to provision nodes across multiple clouds using VPN tunnels against the front-end node of the cluster to provide connectivity across the nodes \cite{Calatrava2016}. The solution proposed in this paper extends EC3 architecture, enhancing network connectivity and enabling the creation of advanced VPN network topologies.

There are also some services and tools for specific cloud providers. For example, Azure CycleCloud \cite{AzureCycleCloud} supports HPC bursts with some existing HPC cluster by delegating extra workload to nodes provisioned from the cloud. Also, AWS ParallelCluster \cite{AWSParallelCluster} is an open-source tool to provision virtual auto-scaled virtual clusters supporting different batch systems in AWS. Similarly, AWS Batch \cite{AWSBatch} is a managed service that allows to deploy virtual clusters that feature scale-to-zero, thus being able to submit a group of jobs to a managed queue that automatically provisions the required virtual machines. These products are only able to provision resources from their own cloud platforms. Finally, Alces Flight \cite{ALCESFlight} supports multiple cloud back-ends (AWS, IBM Softlayer and OpenStack) but does not offer the possibility to deploy clusters that span across multiple cloud providers.

Also, a cluster management tool such as as Kubernetes, combined with tools like Istio \cite{Istio} or Submariner \cite{Submariner}, provides the ability to create a cluster spanning the different cloud sites where users can deploy their applications in a multi-cloud environment. However, this solution is restricted to Kubernetes applications and, therefore, cannot be extended for other cluster management tools. Our solution is more generic since it allows the user to deploy not only Kubernetes clusters, but also other kind of clusters (SLURM, Mesos, Nomad, etc.), where the type of clusters supported can be increased developing the corresponding plugins to the elasticity management framework provided.


As far as networking in hybrid setups is concerned, there are major tools to enable software defined networking (SDN) widely available. The concept of SDN is supported by networking element vendors who provide not only the equipment in the traffic flow plane but also comprehensive management tools for the SDN control plane. An example of this is the Open Networking Environment \cite{CiscoONE} by Cisco Systems. Typically, such solutions are procured, deployed and operated organization-wide.

The same purpose is also pursued by the open source community. The most prominent open-source tool for SDN is  Open vSwitch \cite{OpenvSwitch}. It can serve as a fully virtual networking element or provide a control layer for HW-based switches. It is supported by various data center virtualization frameworks and cloud management platforms such as OpenStack.

Other TOSCA-based orchestration tools such as Cloudify \cite{Cloudify} offer the possibility to orchestrate the creation of network elements (routers, subnets, load balancers, etc.) inside a single cloud provider. However, they do not have the concept of a virtual router enabling the connection of a set of networks in different cloud providers, thus enabling the application to connect with other nodes of the topology (deployed in different cloud providers) as if they were in the same LAN.

Overall, tools for SDN orchestration and operation within institutional boundaries are readily available. For the purpose of hybrid cluster deployment in clouds, however, they are mostly not suited. Firstly, they are typically a part of the organization's underlying technology substrate, and as such their control is never exposed to general users, and certainly not to federated users coming from foreign institutes. Utilizing state-of-the art SDN tools in user space is mostly out of the question. Consequently, the need arises to design a solution for hybrid clusters with a tool that could be fully deployed and controlled in user space, reusing existing, widely available and well proven technologies.

As identified in the revision of the state of the art, there exist several tools and managed services for the deployment of virtual elastic clusters in several cloud platforms. However, most of the solutions rely on a single provider and those that feature multi-clouds do not offer an approach to provision elastic clusters that span multiple cloud providers using techniques to dynamically create virtual networks on which to deploy the virtual routing components required to achieve effective communication among the nodes.

Therefore, the main contribution of this paper to the state of the art lies in the definition of an architecture together with a software implementation that integrates several open-source developments to achieve hybrid virtual elastic clusters across multiple clouds, focusing on network connectivity and security, with L2 network creation and configuration of advanced VPN topologies.

\section{Architecture Design and Components for Hybrid Virtual Clusters\label{sec:architecture}}

This section describes the main open-source components integrated in the architectural solution to deploy hybrid virtual clusters. For each component, a brief description is provided together with the role played by the component in the overall architecture.

\subsection{Overall architecture}

\begin{figure}
\centering
\includegraphics[width=0.75\columnwidth]{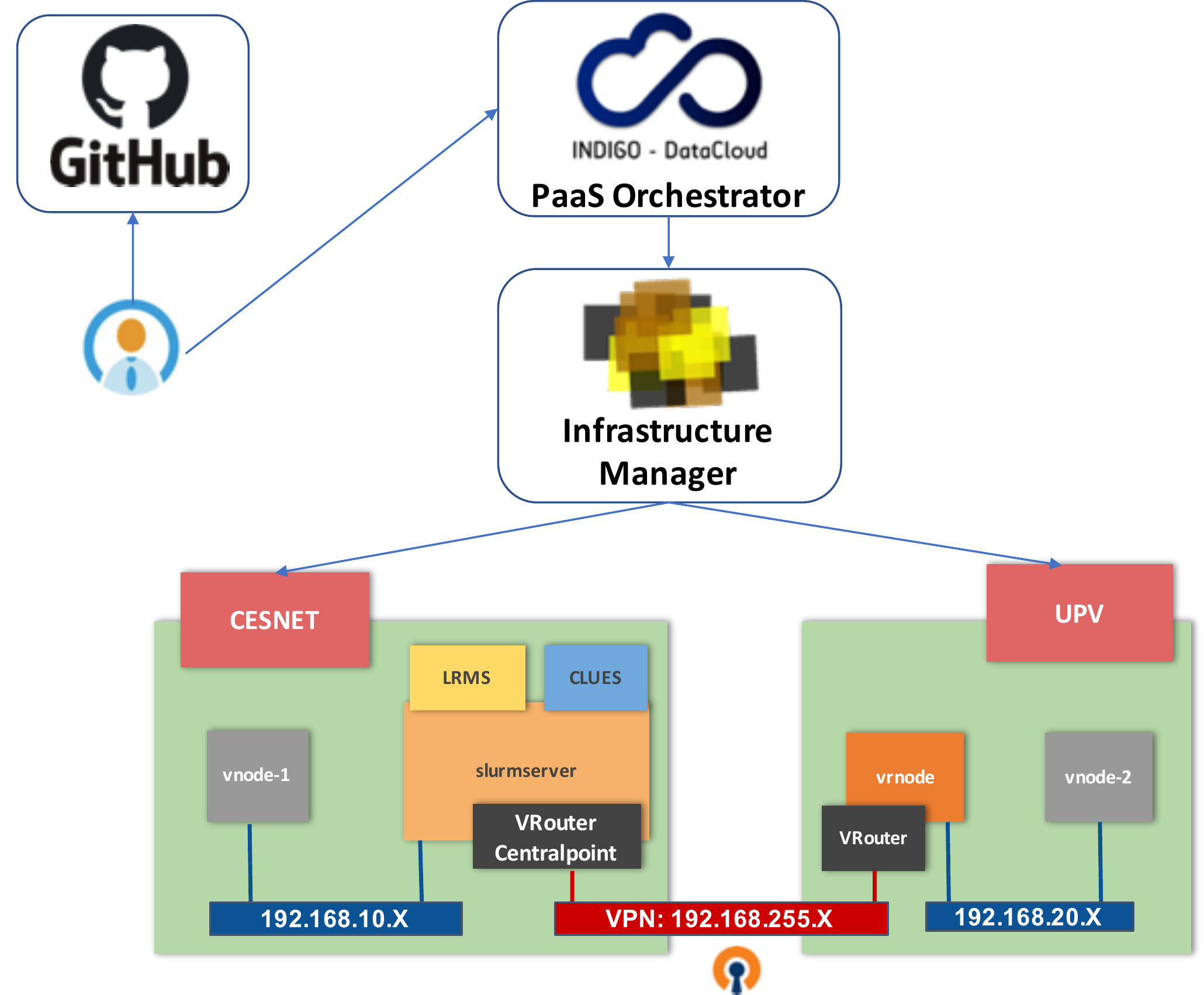}
\caption{\label{fig:arch-l2} Architecture to deploy hybrid virtual elastic clusters across multiple cloud sites (exemplified for two specific cloud sites).}
\end{figure}

Figure \ref{fig:arch-l2} describes the architecture employed to support the deployment of hybrid virtual elastic clusters across sites. In this approach, the front-end node of the cluster (the only VM that will require a public IP) will also act as the vRouter Central Point (CP) in order to avoid requiring an additional VM with a public IP. The rest of VMs hosted in the same provider will not require the installation of any vRouter component, they will only require setting the front-end as the network gateway through DHCP. Then, in each different cloud provider where the cluster's working nodes will be deployed, an extra VM will also be deployed and configured as the vRouter node. It will be configured to route the traffic through the vRouter CP installed in the front-end node. Finally, the rest of VMs deployed on each site will be configured to set the vRouter node deployed in its site as the network gateway. 

The deployment flow starts with the user choosing an existing TOSCA template from a curated repository publicly available in GitHub\footnote{indigo-dc/tosca-templates: \url{https://github.com/indigo-dc/tosca-templates}}.  The template describes an application architecture to be deployed in the cloud. Previous experience by the authors in the area of using TOSCA to describe elastic clusters demonstrated the success in adopting this language also to describe hybrid virtual clusters, as described in the works by Caballer et al. \cite{Caballer2017,Caballer2017tbo}.

Once the TOSCA template is submitted to the PaaS Orchestrator, this component is in charge of choosing the most appropriate sites on which to provision the nodes, depending on the Service Level Agreements (SLAs) available between the user and the cloud sites and considering their monitored availability. Once selected, it delegates on the Infrastructure Manager (IM) to provision from the different cloud providers to create all the virtual resources, including virtual networks to accommodate them. 

The first step is network creation. These private networks are specifically created in each cloud, assuring security by means of network isolation. Afterwards, the IM creates the requested VMs in each of the providers' clouds, attaching them to the newly created networks. Furthermore, the IM will enable SSH reverse tunnels among all the VMs and the front-end node, thus enabling the use of this node as an Ansible control node to configure all the involved VMs without requiring them to have public IP addresses. Finally, the IM will perform the contextualization steps where the VMs will be configured using DevOps tools to create the virtual infrastructure depicted in Figure \ref{fig:arch-l2}. 

\subsection{INDIGO PaaS Orchestrator}

The PaaS Orchestrator allows to coordinate the provisioning of virtualized compute and storage resources on Cloud Management Frameworks, both private and public (like OpenStack, OpenNebula, AWS, etc.), and the deployment of dockerized services and jobs on Mesos clusters. It receives the deployment requests, expressed through TOSCA templates, and deploys them on the best available cloud site. In order to select the best site, the Orchestrator implements a complex workflow: it gathers information about the SLA signed by the providers and monitoring data about the availability of the compute and storage resources.

\begin{figure}
\centering
\includegraphics[scale=0.40, angle=270]{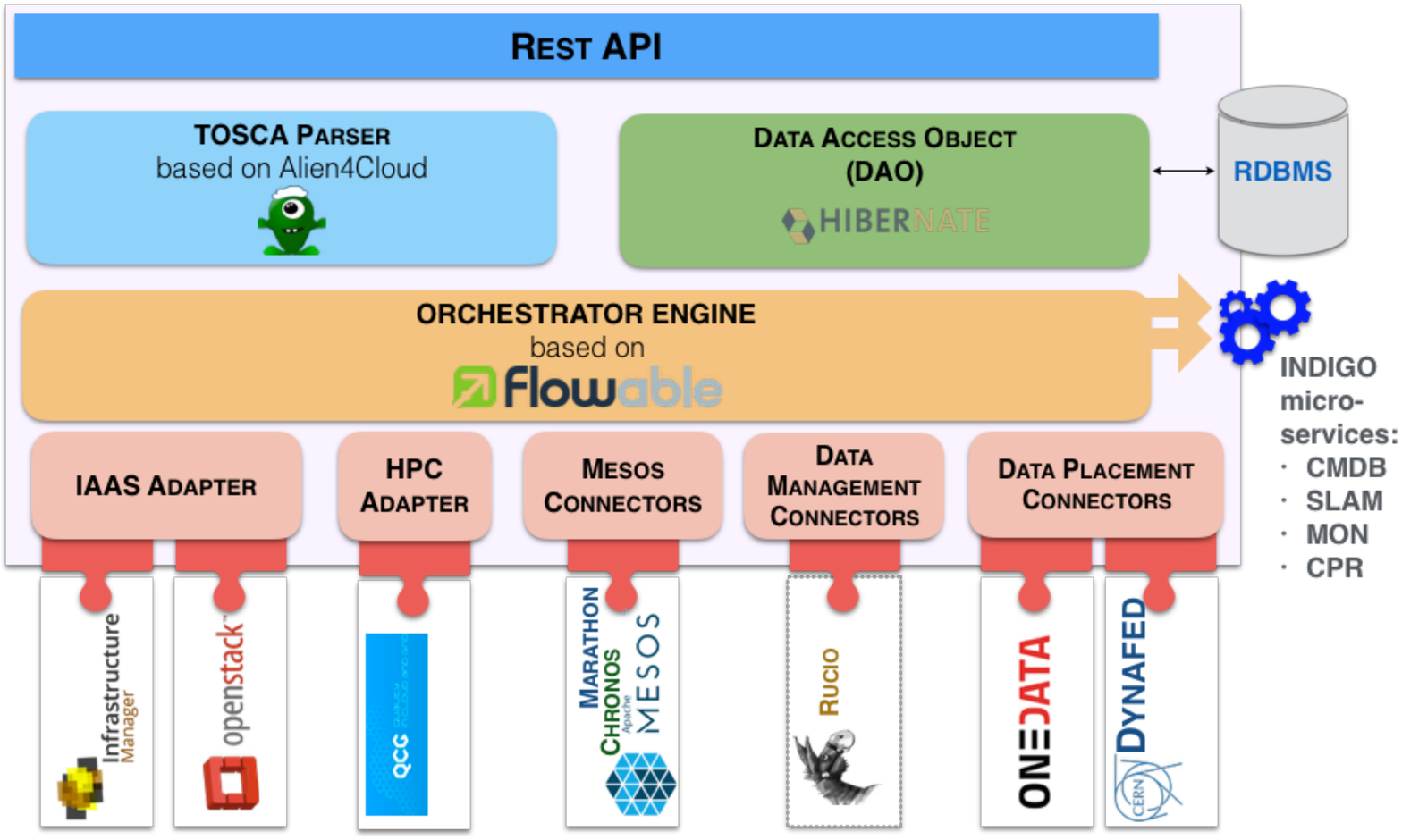}
\caption{\label{fig:orchestrator-arch} PaaS Orchestrator high-level architecture.}
\end{figure}

The Orchestrator architecture (Figure~\ref{fig:orchestrator-arch}) is modular and based on plugins: depending on the deployment requirements, specified in the TOSCA template, the proper adapter/connector is automatically activated:
\begin{itemize}
\item IaaS adapters, in charge of managing the interaction with the Cloud Management Frameworks; this interaction can be performed using directly the Heat Orchestration service for deployments on top of OpenStack clouds or delegating to the Infrastructure Manager.  
\item Mesos connectors, implementing the interfaces that abstract
the interaction with the Mesos frameworks: Marathon\footnote{Marathon: \url{https://mesosphere.github.io/marathon/}} for managing containerized long-running services and Chronos\footnote{Chronos: \url{https://mesos.github.io/chronos/}} for submitting containerized batch-like jobs. 
\item HPC adapter, implementing the interfaces to interact with the
HPC services; the interaction is based on REST APIs provided by a gateway hosted by the HPC site, using QCG Computing APIs\footnote{QCG Computing: \url{http://www.qoscosgrid.org/trac/qcg-computing}}.
\item Data Management connectors, implementing the interfaces to interact with data orchestration systems in order to coordinate data movement and replication requested by the user. The reference implementation is based on the Rucio\footnote{Rucio: \url{https://rucio.github.io/index.html}} system \cite{rucio2019}.  
\item Data placement connectors, in charge of interacting with the storage services in order to get information about the location of user data; this information is used by the Orchestrator to perform data-aware scheduling: if requested by the user, the Orchestrator will try to match compute resources with data availability in order to start the processing near the data.
\end{itemize}

\begin{figure}
\centering
\includegraphics[width=0.9\columnwidth]{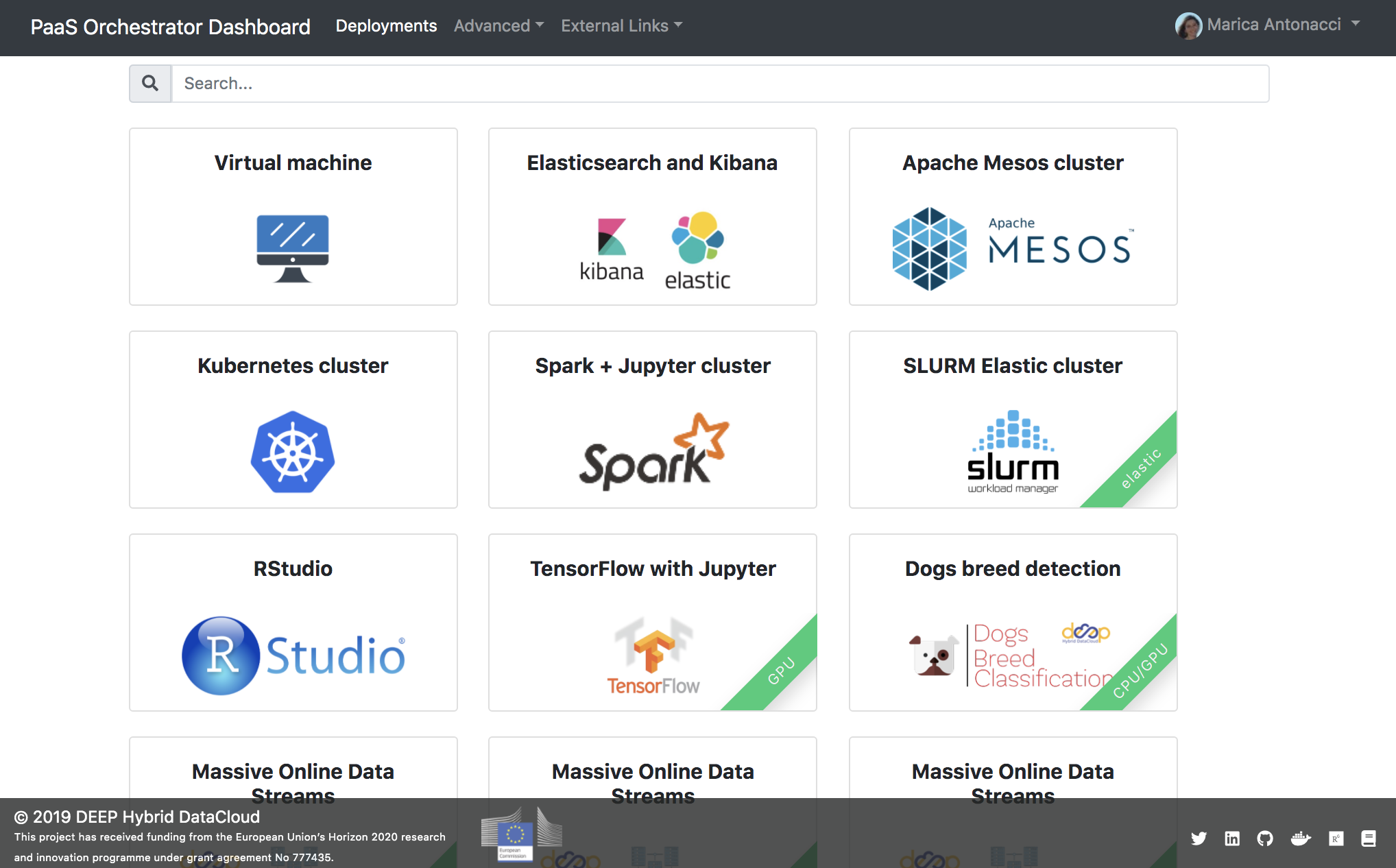}
\caption{\label{fig:orchestrator-dashboard} Orchestrator dashboard.}
\end{figure}

The users can interact with the PaaS Orchestrator via its REST API, also using the \textit{orchent} command-line interface\footnote{\textit{Orchent}: \url{http://github.com/indigo-dc/orchent}} and, finally, via the web-based graphical dashboard, as shown in Figure \ref{fig:orchestrator-dashboard}.

\subsection{Infrastructure Manager (IM)}

\begin{figure}
\centering
\includegraphics[width=0.8\columnwidth]{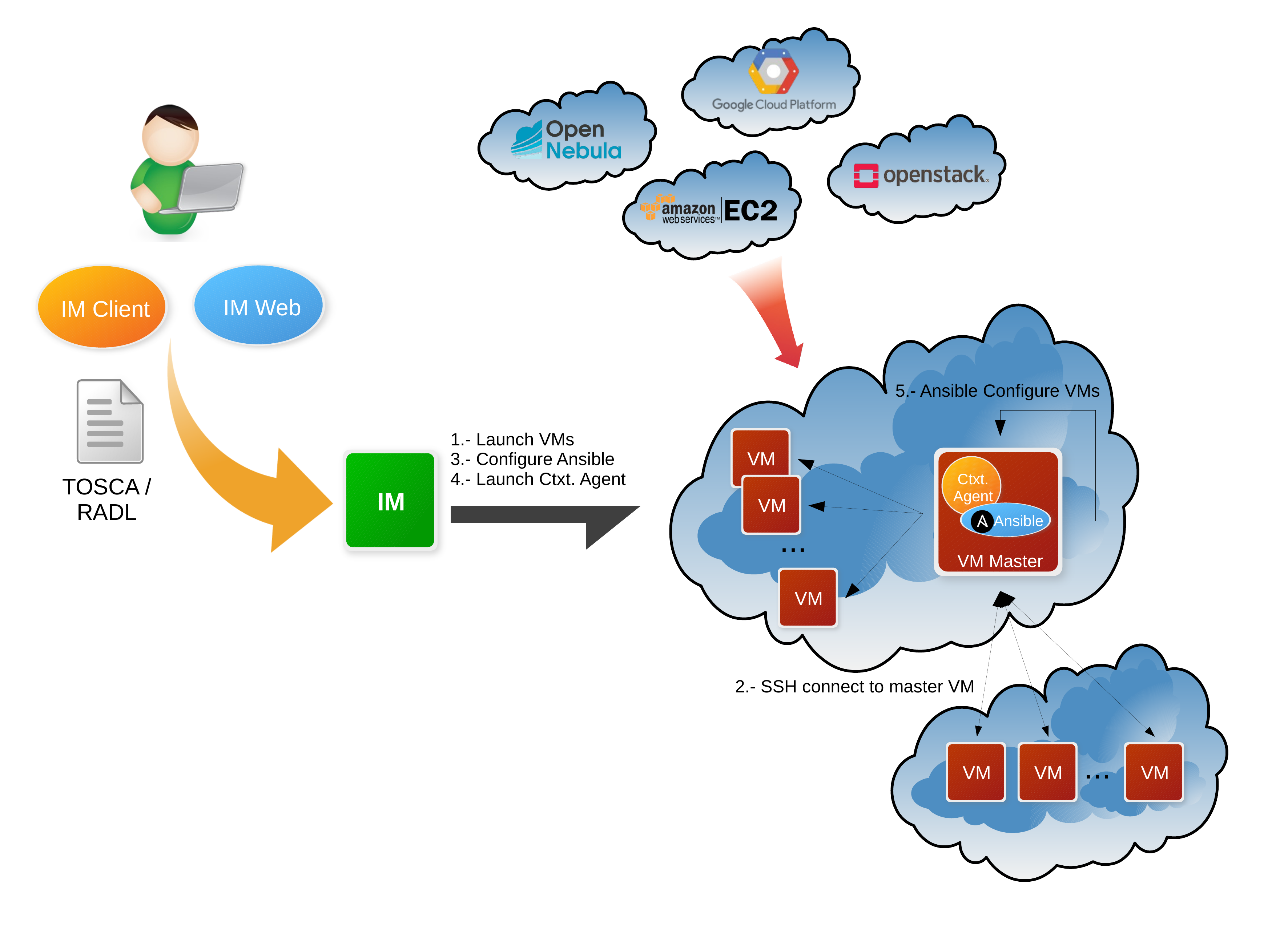}
\caption{\label{fig:im-arch} Infrastructure Manager diagram.}
\end{figure}

The infrastructure Manager (IM)\footnote{Infrastructure Manager: \url{www.grycap.upv.es/im}} \cite{Caballer2015} is a cloud orchestration tool that deploys complex and customized virtual infrastructures on IaaS cloud deployments (such as AWS, Azure, OpenStack, OpenNebula, etc.), as shown in Figure~\ref{fig:im-arch}. It performs automatic deployment of applications described as code, using the OASIS TOSCA (Topology and Orchestration Specification for Cloud Applications) \cite{TOSCA-YAML} and its own description language (Resource and Application Description Language -- RADL). It automates the deployment, configuration, software installation, monitoring and update of the virtual infrastructures. It supports APIs from a large number of virtual platforms, making user applications cloud-agnostic. In addition it integrates a contextualization system (based on Ansible DevOps tool) to install and configure all the user required applications providing the user with a fully functional infrastructure. The IM internally uses Apache Libcloud\footnote{Apache Libcloud: \url{https://libcloud.apache.org/}} in several connectors to interact with the corresponding cloud back-end.

The IM enables the deployment of virtual infrastructures on top of on-premises, public, federated clouds and container orchestration platforms, thus enabling the deployment of virtual hybrid infrastructures which span across multiple providers. Furthermore it uses an SSH reverse approach allowing Ansible to configure all the VMs of an infrastructure from a single VM, known as ``master'', thus requiring only one public IP per infrastructure.

Functionality can be accessed via a CLI, a web GUI, a REST API and a XML-RPC API. It can be deployed in a highly available mode with a set of load-balanced replicated instances in order to cope with a large pool of users. The IM is being used in this architecture to provide multi-cloud provisioning and configuration of the Virtual Machines that will be the nodes of the virtual cluster.

\subsection{CLUES}

CLUES (CLuster Elasticity System)\footnote{CLUES: \url{https://github.com/grycap/clues}} \cite{DeAlfonso2013,DeAlfonso2018} is a general elasticity framework that can be applied both to power on/off nodes of a physical cluster or to provision/terminate virtual nodes from a cloud provider. It supports a wide variety of plugins to introduce elasticity in several LRMS (Local Resource Management Systems) such as HTCondor, SGE, SLURM; container orchestration platforms such as Kubernetes, Mesos and Nomad and on-premises cloud management frameworks such as OpenNebula.

CLUES is being used in this work as the elasticity component employed in the provisioned clusters by the Infrastructure Manager in order to monitor the job queue of the system and decide whether additional nodes should be provisioned or terminated, depending on a set of user-configurable policies. It provides comprehensive reporting and has been successfully used and extended across multiple large-scale European projects such as INDIGO-DataCloud\footnote{INDIGO-DataCloud: \url{www.indigo-datacloud.eu}} and DEEP Hybrid-DataCloud\footnote{DEEP Hybrid-DataCloud: \url{https://deep-hybrid-datacloud.eu/}}.

\subsection{INDIGO Virtual Router}
\label{sec:indigo-virtual-router}

The single most important requirement in virtual networking in the context of hybrid virtual cluster deployments is the ability to support geographically distributed hybrid clusters, wherein different nodes can be located in different sites.
This requirement translates into the need for multi-site spanning private networks providing adequate security and connectivity to virtual machines contained within.

\subsubsection{Motivation}

Designing inter-site virtual networks as private, isolated infrastructures with tightly controlled outside access has multiple reasons. One or more of the following apply in many hybrid deployment scenarios considered and analysed in the framework of the DEEP Hybrid-DataCloud (DEEP) project.

\begin{description}
\item[Fire and forget] The community developing tools for academic hybrid clouds is currently striving to cater to the needs of the so-called ``long tail of science'', entailing small user groups with little manpower to spend on provisioning and maintaining their computing tools. This puts extra emphasis on deployment isolation, since a deployed cluster -- once verified as operational -- tends to get very little attention in terms of maintenance and updates (security or otherwise). As such, it must be tightly isolated from the outside network to prevent exploitation of missing updates, yet full visibility among the nodes of the cluster must be preserved.
\item[Legacy tools] Somewhat similar to the previous point, rather than through aging, cluster deployments are sometimes obsolete by design since the time of their very deployment. Virtual installations are a viable way of continuing to run outdated legacy software, which is a sorely needed capability to assure comparability and reproducibility of scientific results. Again, strict isolation and encapsulation into a private network is a necessity. On top of that, it is highly beneficial for such private network to support Internet Protocol version~4 (IPv4) since many legacy tools require it. Also, there is a disparate degree of adoption of IPv6 for different countries, which turns IPv4 addresses into a scarce network resource at least in some geographical areas \cite{GoogleIPv6}.
\item[Real world resemblance] Cluster software management tools may occasionally employ ``automagic'' algorithms for self-configuration, which assume physical network components. Hence it is also advantageous to make the purpose-built private virtual network resemble a physical one in terms of components and their purpose.
\end{description}

\subsubsection[Design Assumptions]{Design Assumptions}
\label{sec:vrouter-design-assumptions}

The design of the Virtual Router component,\footnote{INDIGO Virtual Router: \url{https://github.com/indigo-dc/ansible-role-indigovr}} originally conceived within the INDIGO-DataCloud project, is based on several key assumptions:

\begin{itemize}
    \item Cluster nodes (whole clusters) need to be located in private networks primarily for security reasons. In case of hybrid deployments such a private network must span all participating sites and comprise all the deployed nodes while ``visibility'' within the network remains unhampered to facilitate auto-discovery and communication across the cluster.
    \item There is negligible chance that the orchestration layer will have control of all intermediate network components across all participating cloud sites, therefore an overlay network is required to provide for a private connection among geographically distant sites.
    \item Some cluster nodes may not allow changes in their internal configuration -- typically those based on custom, user-supplied images whose contents cannot be modified by the orchestrating component. These may come simply as pre-prepared user-produced binary images with no \texttt{root} access, or even as images of ``alien'' operating systems such as MS~Windows, which the orchestration layer does not know how to modify. Hence their networking behavior should be configurable with DHCP and no additional intervention should be required. Note that this ``black-box-like'' behavior is not the case in the particular example presented in this paper's Section~\ref{sec:use-cases}, where the IM actually has control of the internals of individual nodes, but it is a regular occurrence the authors observe while supporting user communities in HTC clouds.
    \item It is impossible (or at least difficult) to rely on a custom appliance image being available in all participating cloud sites. Therefore, the networking component cannot take the form of a binary image but rather that of a ``recipe'' deployable and configurable on standard distribution images (e.g., latest Debian, latest Ubuntu) available universally in all IaaS clouds.
   \item It is likewise impossible (or at least difficult) to rely on solutions bundled with popular cluster software -- such as the \textit{OVN4NFV-K8S-Plugin} for Kubernetes~\cite{OVN4NFVK8s} -- since that would make every cluster deployment depend on Kubernetes or similar heavy-weight cluster stack, limiting the generic nature of cloud VMs.
    \item Finally it may be difficult to deploy the virtual router in a container. While some of its core components (for example a VPN client or server) may lend themselves to containerization quite readily, in general a routing appliance (especially to fulfill its role as a default gateway in a local area network) needs to take over such a large portion of its host system's networking that, firstly, the host system would have to delegate its core networking functions to the container and, secondly, the configuration of the router would have to reside in two locations, i.e., the host system and the containerized system.
\end{itemize}

\subsubsection{vRouter Implementation}

\begin{figure}[bth!]
\hfill\includegraphics[width=.8\textwidth]{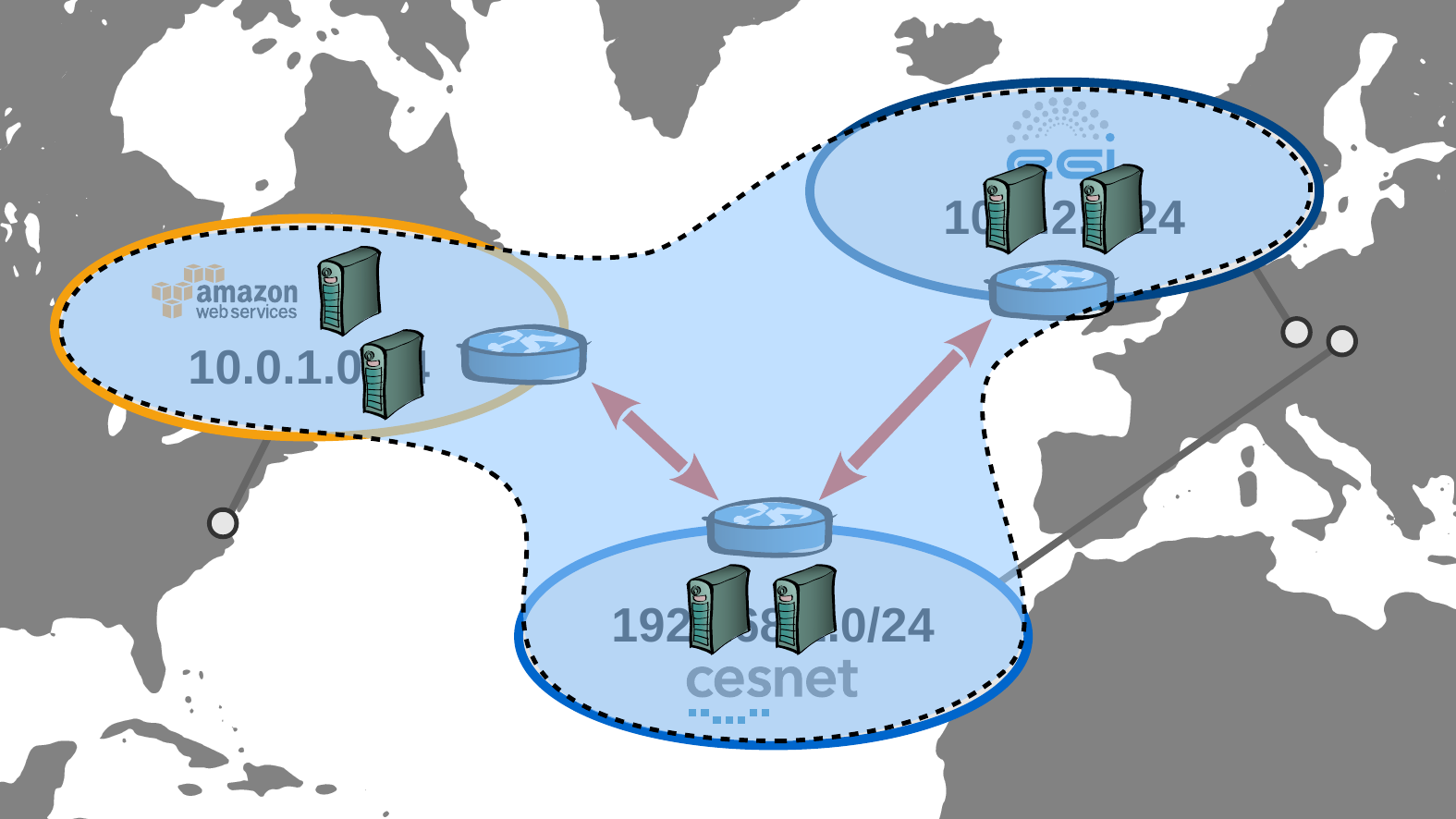}\hfill\null
\caption{A demonstration use case for a hybrid cluster deployment with the INDIGO Virtual Router. A private network spans all participating sites. There is a private virtual network allocated in each cloud, and each has an instance of the vRouter; together, these vRouter instances form the overlay network.}
\label{fig:vrouter-demo-use-case}
\end{figure}

Figure~\ref{fig:vrouter-demo-use-case} shows the simplest use case of cluster deployment with the virtual router (vRouter). There is a virtual private network set up dynamically, specifically for the given deployment in each participating site. Each such network holds an instance of the vRouter. One of the vRouters acts as a Central Point [of the star topology] while the others set up a point-to-point VPN connection to it, routing traffic from their local network through to the central point and onward.

\begin{figure}
\hfill\includegraphics[width=.8\textwidth]{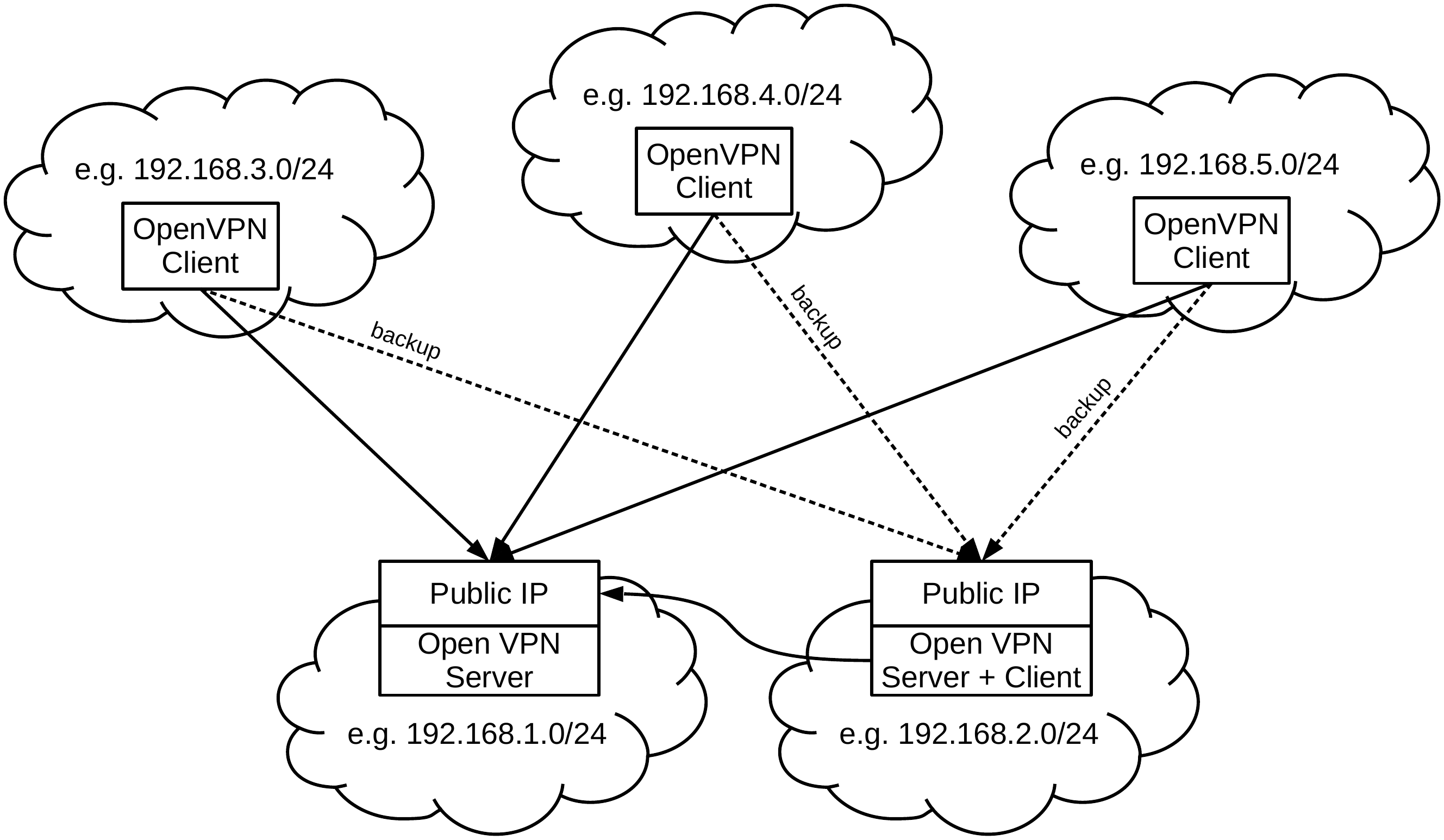}\hfill\null
\caption{An illustration of a redundant star topology with five private networks and five vRouters, out of which two have the role of central point. The remaining vRouters are aware of both central points but would only use their connection to the backup central point (right) if connection to the primary one (left) was lost.}
\label{fig:vrouter-redundancy}
\end{figure}

Each vRouter is a simple virtual machine (VM), which runs primarily an OpenVPN component (client or server). The appliance may be deployed in two roles:
\begin{description}
\item[vRouter] routes traffic between nodes in the local private network and remote sites participating in the deployment. It may also comprise a DHCP server to advertise desired network configuration to nearby cluster nodes. This capability, however, is used typically only if the local cloud middleware cannot perform custom DHCP configuration by itself.
\item[Central Point] is a designated vRouter. There must be at least one in each deployment. It accepts VPN connections established by individual vRouters. It has the intended network configuration of the deployment pre-configured and as clients connect it assigns subnet address ranges and other configurations. It is the only node within the whole deployment requiring a public IP address. The topology may contain multiple Central Points (Figure~\ref{fig:vrouter-redundancy}) but currently an additional central point will only be used as a hot backup.
\end{description}

\subsubsection{Support for Stand-alone nodes}

Although not originally assumed by the INDIGO Virtual Router design, it turned out that there was a demand for supporting ``stand-alone'' nodes, be it cluster nodes or other machines the user wishes to connect to the setup. Such stand-alone nodes exist typically in public or shared networks, which cannot be fully dedicated to the given hybrid cluster deployment. That makes it impossible for a vRouter instance to take control of the whole subnet. As a solution, the concept of a stand-alone node has been introduced.

\begin{figure}
\hfill\includegraphics[width=.8\textwidth]{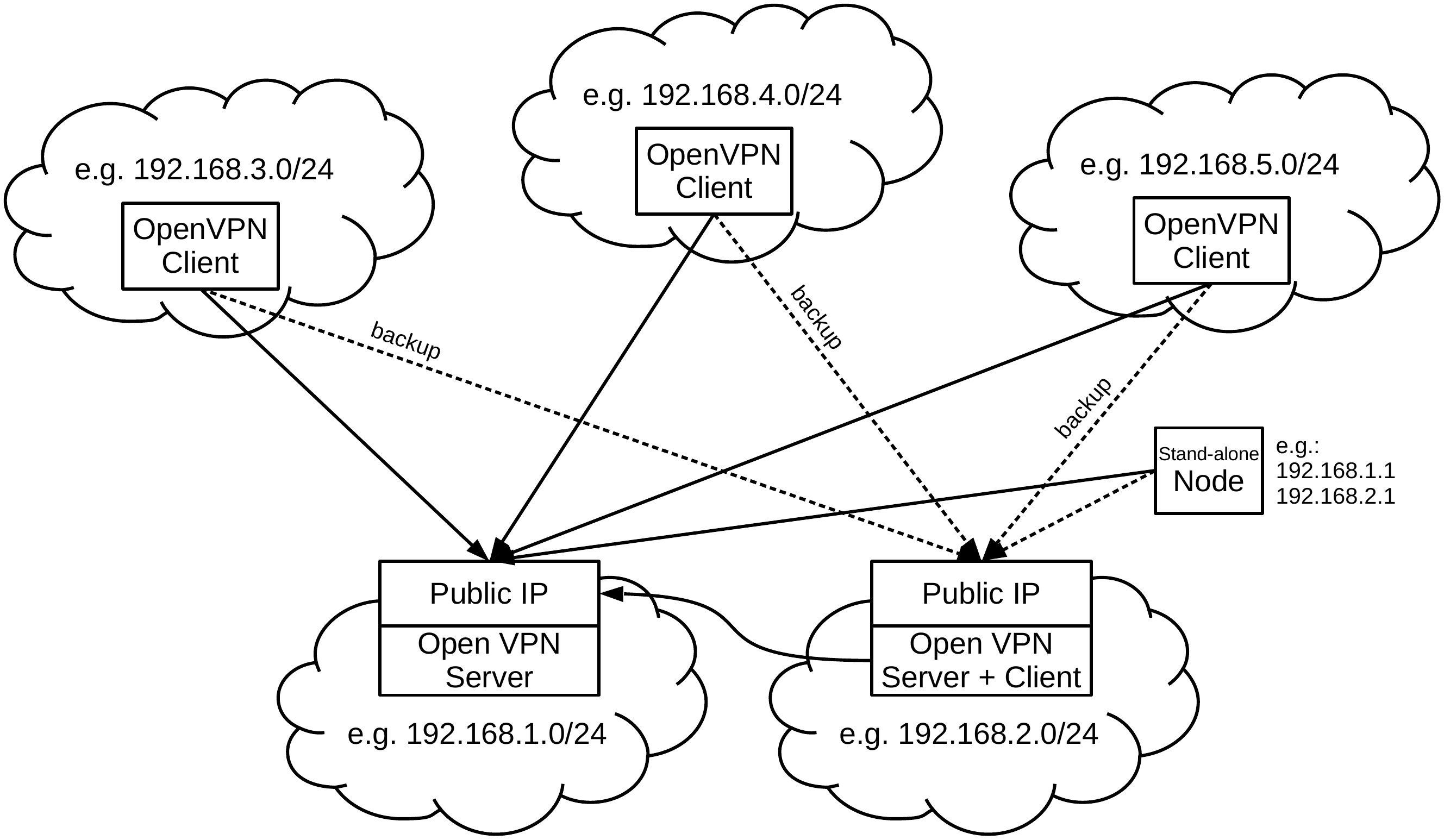}\hfill\null
\caption{A complex virtual private network topology with a stand-alone node located outside any of the vRouter-controlled local virtual networks}
\label{fig:vrouter-stand-alone}
\end{figure}

Stand-alone nodes will typically appear for one of two reasons:
\begin{enumerate}
    \item The participating cloud does not support creation of virtual private networks, or will not otherwise allow the vRouter to take control of the private network, yet the orchestration layer has decided to place cluster node(s) in that site.
    \item The user wishes to connect a pre-existing machine (e.g., their own workstation) to the deployment.
\end{enumerate}
Should either be the case, a VPN client can be installed directly on the stand-alone node, and a VPN connection established directly with the central point (Figure~\ref{fig:vrouter-stand-alone}). This, however, breaks the third assumption made in Section~\ref{sec:vrouter-design-assumptions} as the orchestration layer must now be able to directly install and configure components (VPN clients) on the nodes. This precludes the use of ``black box'' images for setting up cluster nodes, but as a trade-off allows for hybrid cluster deployment in sites that will not expose sufficient control capabilities for networking resources.

\subsubsection{Overlay Network Security}

The overlay inter-site network deployed in this manner can be considered fully secured. The individual local networks in participating cloud sites are private, established on purpose for the given deployment (or, at worst, assigned from a pool of pre-configured networks). Inter-site connections between vRouter instances and possible stand-alone nodes are built with OpenVPN and as such they are also secure.

OpenVPN uses X.509 certificates for its clients to authenticate to servers. With client identities (certificate subjects) pre-registered at the server (CP) it is even possible to assign a static configuration to each node. That makes it possible for the orchestration layer to pre-determine which client vRouter will be assigned which subnet.

The decision to rely on host certificates to establish trust between vRouter instances led to the need for an ability to generate said certificates, distribute them and register their subjects with the CP. Third-party solutions were considered but finally rejected because, at that point, there was no other use case requiring a separate, stand-alone trust storage solution. OpenVPN comes bundled with the \textit{Easy RSA} tool, which could be easily reused for this purpose. Alternatively, the OpenSSL suite also provides tools to manage certificates. Certificates are generated at the CP, and Infrastructure Manager's existing call-back function is used to retrieve the generated client certificates or even request additional ones in more dynamic use cases. This makes it possible for the orchestration layer to distribute secrets and establish complete trust between networking elements within the deployment relying solely on pre-existing tools already available in the DEEP stack, without the need to introduce additional tools into the mix.

\subsubsection{Performance-Security Tradeoff}

An encrypted OpenVPN connection between local networks may prove to present a bottleneck, especially should the hybrid cluster deployment require extensive inter-node communication. This may be perceived as a problem if inter-node communication is already encrypted natively by the cluster software, making another layer of encryption by OpenVPN superfluous. 

This particular issue is easily tackled by configuring OpenVPN to use a less costly encryption algorithm or even no encryption at all. That would be perfectly adequate for clusters based on software that is already prepared for working over the public Internet. However it is not the default option because most relevant deployments will assume they may rely on the safety of a local private network.

In the future it should be also possible to alleviate the burden imposed especially on the central point by enabling dynamic identification of shorter network paths within the deployed topology.

\section{Use Case: Scalable Batch Computing Across Geographic and Administrative Boundaries \label{sec:use-cases}}

In order to test the suitability of the defined architecture we designed a use case that fully embraces the term hybrid. First, it consists of the deployment of a virtual elastic cluster across two different cloud sites. Second, one will be a public cloud provider and the other will be an on-premises cloud being used in production in a research center. This way, it will be demonstrated that the cluster can span across multiple organizational entities and across different technology stacks.

\begin{figure}
\centering
\includegraphics[scale=0.35]{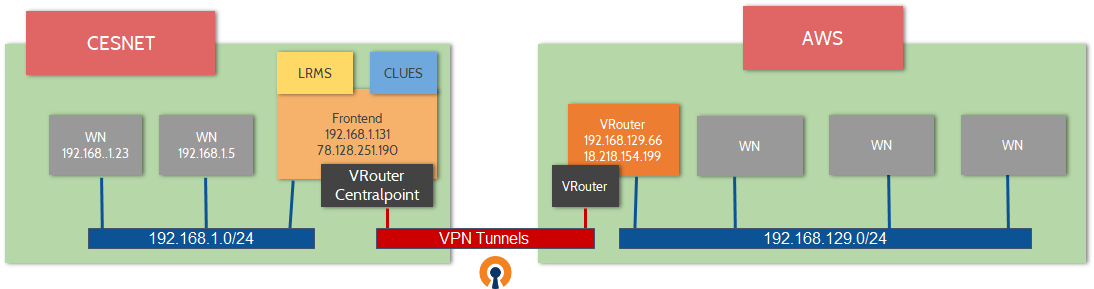}
\caption{\label{fig:hybrid-setting} Deployed virtual hybrid infrastructure across cloud sites.}
\end{figure}

\subsection{Use Case Description\label{sec:description}}

These are the cloud sites involved in the use case:
\begin{description}
\item[MetaCentrum Cloud (MCC)] Operated by CESNET, MCC is based on OpenStack as its Cloud Management Platform and supports federated authentication through OIDC. The site was used in prototyping and deployment testing hybrid cluster deployments with vRouter-based virtual networks.

\item[Amazon Web Services] The largest public cloud provider to-date. In particular, the region \textit{us-east-2} (Ohio) was used. VPC (Virtual Private Cloud) together with dynamically created private networks has been used to accommodate the execution of Virtual Machines. The instance type t2.medium (2\,vCPUS, 4\,GBs of RAM) was chosen since it provides an adequate compromise between hourly price (billed by the second) and performance required to perform the execution of the jobs. A plain Ubuntu 16.04 image was used in both cases as the base image of the VMs.

\end{description}

The hybrid virtual elastic cluster creation has been made using the Orchestrator dashboard shown in Figure \ref{fig:orchestrator-dashboard}. The user selects the ``SLURM Elastic cluster" option and fills in the required fields (maximum number of working nodes, CPU and amount of memory of the nodes, etc.). Finally, the dashboard submits the deployment to the PaaS Orchestrator, in order to start the deployment process.

The deployment of the hybrid virtual elastic cluster proceeded according to the following steps (corresponding to the scenario shown in Figure \ref{fig:hybrid-setting}):

\begin{enumerate}
    \item Only the front-end node (FE) of the cluster is initially deployed at CESNET, and configured with CLUES in order to detect when pending jobs are queued up at the LRMS (SLURM was used). The FE node does not execute jobs since it is in charge of executing the control plane (shared NFS volume across all the nodes, vRouter, SLURM, etc.)
    
    \item Up to two additional nodes are deployed at CESNET in order to configure the first part of the cluster. This represents a situation where existing quotas at the on-premises cloud prevent a user from deploying additional nodes.
    
    \item Jobs are submitted as workload to the virtual cluster. This causes the allocation of jobs to be executed in any of the two working nodes (WN) allocated in CESNET.
    
    \item Once all the executing slots are filled, CLUES triggers the provisioning of additional WNs through the PaaS Orchestrator, which  provisions the nodes from AWS.
    
    \item CLUES is responsible for monitoring the state of the WNs and the LRMS to decide on allocating additional WNs or terminating them, when no pending jobs are available to be executed.
    
\end{enumerate}

These executions have used the Audio Classifier model \cite{DEEP-Audio-Model} from the DEEP Open Catalog \cite{DEEP-Open-Catalog}. This is a Deep Learning model to perform audio classification that was initially pre-trained with 527 high-level classes from Google's AudioSet dataset \cite{AudioSet}, packaged as Docker images. To easily execute the audio classifier in the virtual cluster, as jobs sent through the LRMS, the udocker tool \cite{Gomes2018} was used in order to run Docker containers in user space. 

To produce a real-world workload we selected a subset of the Urban Sound Datasets \cite{Salamon2014} that consisted of the first 4 audio folders with 3,676 audio files to process and a total of 2.8\,GB.

\begin{figure}[h]
\centering
\includegraphics[width=\columnwidth]{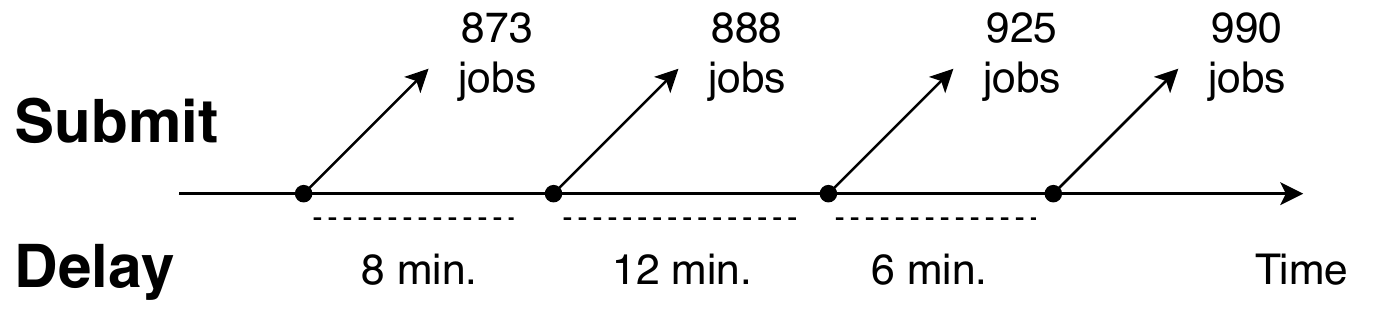}
\caption{\label{fig:timeline} Workload timeline.}
\end{figure}

The workload was split in 4 blocks, including a certain waiting time in between to demonstrate the ability of the elasticity manager to add/remove nodes dynamically based on the system workload. The workload timeline is shown in Figure~\ref{fig:timeline}.

Each processing job includes four steps:
\begin{itemize}
\item Download and install udocker (in case it is not already installed in the WN)
\item Pull the audio classifier Docker image from Docker Hub (in case it is not already available in the WN from a previous pull operation).
\item Create the container (in case it has not been created before).
\item Process the audio file
\end{itemize}

Notice that the first three steps are only carried out once per node, and they took an average total of 4\,min 30\,secs. The audio processing step usually takes about 15 -- 20\,secs.

Each job processes an audio (WAV) file and creates a JSON output file with the labels of the most accurately predicted categories of sounds together with links to both Google and Wikipedia, as shown in the following excerpt:

\begin{lstlisting}
[{'title': '98223-7-4-0.wav', 'labels': ['Jackhammer', 'Speech', 'Vehicle', 'Sawing', 'Engine'], 'probabilities': [0.12856674194335938, 0.12842661142349243, 0.10691297054290771, 0.10328176617622375, 0.09795475006103516], 'labels_info': ['/m/03p19w', '/m/09x0r', '/m/07yv9', '/m/01b82r', '/m/02mk9'], 'links': {'Google Images': ['https://www.google.es/search?tbm=isch&q=Jackhammer', 'https://www.google.es/search?tbm=isch&q=Speech'], 'Wikipedia': ['https://en.wikipedia.org/wiki/Jackhammer', 'https://en.wikipedia.org/wiki/Speech']}}]    
\end{lstlisting}

\subsection{Results and Discussion\label{sec:results}}

\begin{figure}
\centering
\includegraphics[width=\columnwidth]{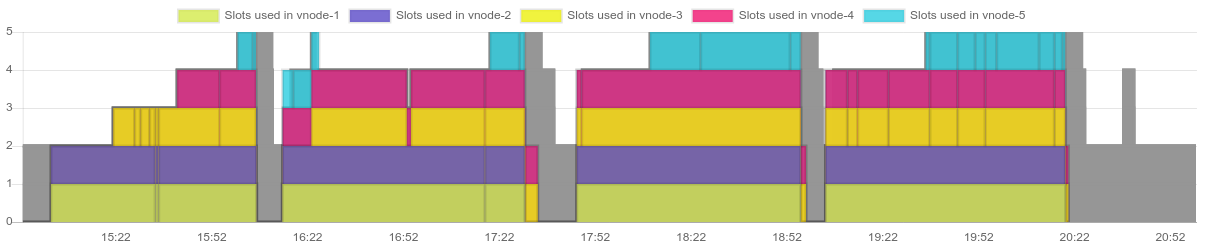}
\caption{\label{fig:cluster-evolution} Cluster usage evolution.}
\end{figure}

Figure \ref{fig:cluster-evolution} shows the usage of all participating nodes during the test. Note that \textit{vnode-1} and \textit{vnode-2} correspond to the WNs initially deployed at CESNET while the remaining WNs correspond to those dynamically deployed in AWS on-demand. It took approximately 19 minutes for AWS nodes to be deployed, configured, and added to the SLURM cluster. 

\begin{figure}
\centering
\includegraphics[width=\columnwidth]{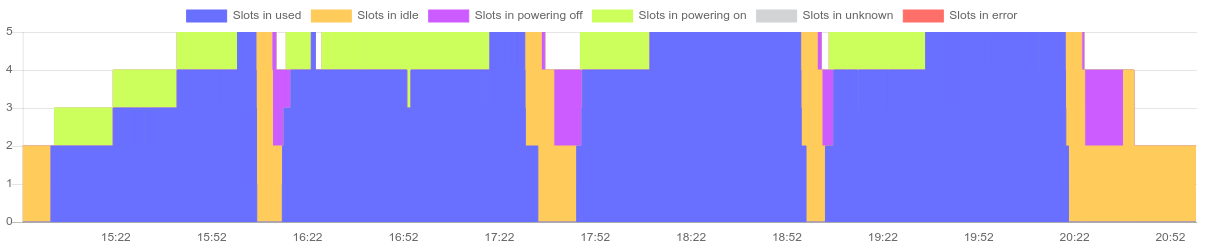}
\caption{\label{fig:node-evolution} Node state evolution.}
\end{figure}

Figure~\ref{fig:node-evolution} shows the states of the nodes. The number of nodes that are used (executing jobs) are shown in blue; green indicates that the nodes are powering on (until they are fully configured and added to the cluster); orange is reserved for idle nodes and purple is used for nodes that are powering off (until they are finally destroyed). It clearly shows the evolution of the cluster's status in time: how the nodes are powered on when the first set of jobs arrives to the cluster (15:00) until all of them are powered on and in use (16:00). These 60 minutes correspond to the time required to power on three nodes (20 minutes each). Then, at 16:05, the first set of jobs is finished and all the nodes become idle. Finally, three of them are selected to power off until the next set of jobs arrives. But the early arrival of new jobs made CLUES cancel the pending power off operations on the used nodes, thus making only vnode-3 actually power off. A problem appeared in this second step: vnode-5 is detected within a few minutes as ``off'' by the SLURM manager and, thus, CLUES marks the node as ``failed'' and powers it off to avoid unnecessary costs by failed VMs. Then, since there are remaining jobs, CLUES powers it on again. This introduces an additional delay until vnode-5 can start processing jobs. Lastly, in the final two steps, the behavior is once again as expected and only ``vnode-5'' is powered off and on, respectively, when the set of jobs ends/starts again.

The total duration of the test was 5 hours and 40 minutes, with a total CPU usage of 20 hours. In particular the total time needed to execute all the jobs was 5 hours and 20 minutes. Twenty extra minutes were required for the additional AWS nodes to power off correctly.

The PaaS Orchestrator workflow engine has a limitation in that it does not allow a deployment to be modified (nodes added or removed) while an update operation is in progress. As a result of this, Figures~\ref{fig:cluster-evolution} and \ref{fig:node-evolution} show the steps of about 20 minutes spent deploying additional nodes (15:00 -- 15:20, 15:20 -- 15:40 and 15:40 -- 16:00).

AWS WNs were executing jobs for 9\,h 42\,m and there were about 5 hours spent in idle or in the power on/off processes. Therefore, 66\% of the paid time of these nodes was used in effective job computation. Notice that the aforementioned limitation in the Orchestrator is causing the nodes to spend a relevant amount of time in the power on/off operations, since multiple node deployments cannot be performed simultaneously.

The total cost of the AWS allocated resources of this test was 0.75\,\$. This cost includes near 15\,CPU hours of the three WNs deployed at AWS (5:31 -- vnode-3, 4:45 -- vnode-4 and 4:25 -- vnode-5) and 6 extra hours of the VRouter instance.

Assuming that there was no possibility of extending the cluster with AWS resources, the workload that required 9\,h and 42 mins of computation would have been distributed among the only two nodes that were available on the CESNET site. Thus, considering that CESNET WNs have similar features to the selected AWS instance types, the test would have required approximately four extra hours to complete all the jobs. Note that the user of the cluster did not need to change anything in their deployment procedure, apart from submitting the jobs and wait for them to finish. Furthermore, all the external network connections were made through secured VPN networks assuring the security and confidentiality of data being transferred among the nodes. Therefore, being able to provision clusters across hybrid cloud deployments allows to pool a significantly larger amount of resources in order to decrease the amount of time required to process high-throughput computing tasks.

\section{Conclusions and Future Work\label{sec:conclusions}}

This paper has introduced an architecture to automatically deploy hybrid virtual infrastructures across multiple cloud platforms. This has been exemplified for the use case of elastic virtual clusters, where these computing entities can span across geographical and administrative boundaries. These clusters pool computing resources from different cloud sites, integrated within a single batch system, and have the ability to deploy additional working nodes and terminate them when no longer needed. 

A real-world workload based on an existing dataset to be inferenced with available Deep Learning models from an open catalog has been adopted as use case to demonstrate the effectiveness of the designed platform. The integration of resources both from an on-premises cloud and from a public cloud provider has proved its benefits for cloud bursting within a single computing entity, which is the virtual cluster. This allows users to be completely abstracted from deciding which cloud sites the resources are actually being provisioned in.

The integration of the virtual router has provided seamless connectivity among the virtual nodes of the cluster, including automated encryption of communications, all dynamically deployed as part of the virtual infrastructure provision.

Future work includes performing large-scale tests involving a wide number of cloud sites in order to determine the bottlenecks of the developed approach. Also, the integration of both CPU and GPU based resources within the same virtual cluster entity pooled from multiple cloud sites and made available to users via different batch queues. Furthermore, optimising the ability to perform parallel provisioning of nodes in the PaaS Orchestrator will reduce the deployment time.

Another set of future work objectives relates to dynamic balancing of inter-cloud virtual network connections.
The private virtual overlay networks based on the INDIGO Virtual Router, as explained in Section~\ref{sec:indigo-virtual-router}, already have the advantage of resembling actual physical ``metropolitan area'' networks (MAN) in their topology. They consist of multiple local networks (analogical to LANs) with routers directing traffic to and from distant networks -- other LANs within the deployment. This is perceived as a benefit since it makes it easy for platform users to understand the topology of their deployment without learning new networking concepts.
What is currently missing from the design, though, is the dynamic identification of the best path for each data frame, which has become the hallmark of IP-based traffic on the Internet. While the vRouter can be configured to recognize and maintain connections to multiple routing counterpoints within the virtual infrastructure, only one such connection is used as a primary route and the others would only serve as ``hot backup'', ready to take over in case the primary connection (or vRouter central point) fails.

Extending the overall configuration of the overlay deployment so that it mimics actual physical networks even in that trait of automatic optimum (shortest) path lookup would be a logical and potentially quite worthwhile step forward.

\section{Acknowledgement}

The work presented in this article has been partially funded by project DEEP Hybrid-DataCloud (grant agreement No 777435). GM and MC would also like to thank the Spanish ``Ministerio de Economía, Industria y Competitividad'' for the project ``BigCLOE'' with reference number TIN2016-79951-R. Computational resources at CESNET, used in the real-world use case, were supplied by the project ``e-Infrastruktura CZ'' (e-INFRA LM2018140) provided within the program Projects of Large Research, Development and Innovations Infrastructures.

\bibliography{bibitems}

\end{document}